\documentclass[12pt]{article}

  \usepackage{amsfonts}
\begin{document}

\begin{center}
{\large \bf Higgsless Electroweak Model and Contraction of Gauge Group }
\end{center}

\begin{center}
N.A.~Gromov \\
Department of Mathematics, Komi Science Center UrD, RAS, \\
Kommunisticheskaya st. 24, Syktyvkar 167982, Russia \\
E-mail: gromov@dm.komisc.ru
\end{center}

\begin{abstract}
A modified formulation  of the Electroweak Model with  3-dimensional spherical geometry in the %matter fields 
 target space is suggested. The {\it free} Lagrangian in the spherical field space along with the standard gauge field Lagrangian form the full Higgsless  Lagrangian of the model, whose second order terms reproduce the same experimentally verified fields with the same masses as the Standard Electroweak Model. 
The vector bosons masses are automatically generated, so there is no need in special mechanism of spontaneous symmetry breaking.

The limiting case of the  modified Higgsless Electroweak Model, which corresponds to the contracted gauge group $SU(2;j)\times U(1)$ is discussed.
%, where $j=\iota$ or $j \rightarrow 0$.
%The  limiting case of the bosonic part of the standard Electroweak Model, which correspond to the contracted gauge group $SU(2;j)\times U(1)$ 
%is discussed.  
Within framework of the limit model 
Z-boson,  electromagnetic and electron fields are interpreted as an external ones with respect to  W-bosons and neutrino fields. The W-bosons and neutrino fields do not effect on these external fields. The masses of the all  particles %of the Electroweak Model 
remain the same, 
%under contraction.
but the field interactions in contracted model are more simple as compared with the standard Electroweak Model.% due to nullification of some terms.
\end{abstract}

%{\bf Keywords}: gauge theory; electroweak model; Higgs boson

%\pacno{12.15--y} 

    %\maketitle
  
\section{Introduction }

The Standard Electroweak Model  based on gauge group $ SU(2)\times U(1)$ gives a good  description of electroweak processes.
% exept for Higgs boson, which has not been experimentally verified up to now. 
One of the unsolved problems is the origin of electroweak symmetry breaking.  
In the standard formulation the scalar field (Higgs boson)  performs this task via Higgs mechanism,  which  generates a mass terms for vector bosons. 
Sufficiently artificial Higgs mechanism  with its imaginary bare mass is a naive relativistic analog of the  phenomenological description of superconductivity \cite{Sh-09}.
However, it is not yet  %has still not been 
experimentally verified whether electroweak symmetry is broken by such a Higgs mechanism, or by something else.
 The emergence  of  large number  Higgsless models  \cite{C-05}--\cite{MT-08} was stimulated by difficulties with Higgs boson. These models are mainly based on extra dimensions of different types or  larger gauge groups. 
A finite electroweak model without a Higs particle which is used  a regularized quantum field theory \cite{E-66},\cite{E-67} was developed in
\cite{MT-08}.

One of the important ingredient of the Standard  Model is the simple group $ SU(2)$. More then fifty years  in physics it is well known 
the notion of group contraction \cite{IW-53}, i.e. limit operation, which transforms, for example, a simple or semisimple group to a non-semisimple one. From general point of view
for better understanding of a physical system it is  useful to  investigate  its properties for limiting values of their physical parameters.
In partucular, for a gauge model one of the similar limiting case corresponds to a  model with contracted gauge group.
The gauge theories for non-semisimple groups which Lie algebras admit invariant non-degenerate metrics was considered in \cite{NW-93},\cite{T-95}.
%The  construction given in   \cite{S-07} is based on  an observation: the underlying group of the Standard  Model can be represented as a semidirect product of $U(1)$ and $SU(2)$.

In the present paper a modified formulation  of the Higgsless Electroweak Model and its limiting case for contracted gauge group 
 $ SU(2;j)\times U(1)$ is regarded.
Firstly we observe that the quadratic form $\phi^\dagger\phi=\phi_1^*\phi_1+\phi_2^*\phi_2=R^2$
of the complex matter field $\phi \in {\bf C}_2$ is invariant with respect to gauge group transformations 
$ SU(2)\times U(1)$ and we can restrict fields on the quadratic form  without the loss of gauge invariance of the model. This quadratic form define three dimensional sphere in four dimensional Euclidean space of the real components of $\phi,$ where the noneuclidean spherical geometry is realized.

Secondly we introduce the {\it free} matter field Lagrangian in this spherical field space, which along with the standard gauge field Lagrangian form the full Higgsless Lagrangian of the model. Its second order terms reproduce the same fields as the Standard Electroweak Model but without the remaining real dynamical Higgs field. The vector bosons masses are automatically generated and are given by the same formulas as in the Standard Electroweak Model, so there is no need in special mechanism of spontaneous symmetry breaking.
The  fermion Lagrangian of the Standard Electroweak Model are modified  by replacing of the  fields $\phi$ with the restricted on the quadratic form fields  in such a way that its second order terms provide the electron mass and neutrino remain massless.

We recall the definition and properties of the contracted group $SU(2;j)$ in Sec. 2. 
In Sec. 3, we step by step modify
the main points of the Electroweak Model for the gauge group $SU(2;j)\times U(1)$.
We find transformation properties %(\ref{eq28}),(\ref{eq29}) 
of gauge and matter fields under contractions. After that we obtain the Lagrangian of the contracted model  from the noncontracted  one  by the substitution of the transformed  fields.   
The limiting case of the modified Higgsless  Electroweak Model is regarded in Sec. 4.  
When contraction parameter tends to zero $j \rightarrow 0$ or takes nilpotent value $j=\iota$ the field space is fibered \cite{Gr-09} in such a way that electromagnetic, Z-boson and electron fields are in the base whereas charged W-bosons and neutrino fields are in the fiber.
Within framework of the limit model the base fields can be interpreted as an external ones with respect to the fiber  fields
  in the sence that the fiber  fields   do not effect on the base fields. 
%In addition  field interactions in contracted model are more simple as compared with the standard Electroweak Model.
The field interactions are simplified under contraction.
Sec. 5 is devoted to the conclusions.

\section{Contracted Special Unitary group  $SU(2;j)$ }

Let us regard  two dimensional complex fibered vector space $\Phi_2(j)$ with one dimensional base $\left\{\phi_1\right\}$
and one dimensional fiber $\left\{\phi_2\right\}$ \cite{Gr-94}. This space has two hermitian forms: first  in the base $\bar{\phi_1}\phi_1=|\phi_1|^2$ and second in the fiber  $\bar{\phi_2}\phi_2=|\phi_2|^2,$ where bar denotes complex conjugation. Both forms can be written by one formula
\begin{equation}
\phi^\dagger(j)\phi(j)=|\phi_1|^2+ j^2|\phi_2|^2,
\label{g1}
\end{equation}  
where $\phi^\dagger(j)=(\bar{\phi_1},j\bar{\phi_2}), $
parameter $j=1, \iota$ and $\iota$ is nilpotent unit $\iota^2=0.$  
For nilpotent unit the following heuristic rules be fulfiled: 
1) division of a real or complex numbers by $\iota$ is not defined, i.e. for a real or complex $a$ the expression $\frac{a}{\iota}$
is defined only for $a=0$, 
2) however identical nilpotent units can be cancelled $\frac{\iota}{\iota}=1.$

 The special unitary group $SU(2;j)$ is defined as a transformation group of $\Phi_2(j)$ which keep invariant the hermitian form (\ref{g1}), i.e.
$$ 
\phi'(j)=
\left(\begin{array}{c}
\phi'_1 \\
j\phi'_2
\end{array} \right)
=\left(\begin{array}{cc}
	\alpha & j\beta   \\
-j\bar{\beta}	 & \bar{\alpha}
\end{array} \right)
\left(\begin{array}{c}
\phi_1 \\
j\phi_2
\end{array} \right)
=u(j)\phi(j), \quad
$$
\begin{equation}
\det u(j)=|\alpha|^2+j^2|\beta|^2=1, \quad  u(j)u^{\dagger}(j)=1.
\label{g3}
\end{equation}  

 The fundamental representation of the one-parameter subgroups of  $SU(2;j)$  are easily obtained
\begin{equation}
u_1(\alpha_1;j)=e^{\alpha_1T_1(j)}=\left(\begin{array}{cc}
	\cos \frac{j\alpha_1}{2} & i\sin \frac{j\alpha_1}{2} \\
i\sin \frac{j\alpha_1}{2}	 & \cos \frac{j\alpha_1}{2}
\end{array} \right),
\label{g4}
\end{equation}  
\begin{equation}
u_2(\alpha_2;j)=e^{\alpha_2T_2(j)}=\left(\begin{array}{cc}
	\cos \frac{j\alpha_2}{2} & \sin \frac{j\alpha_2}{2} \\
-\sin \frac{j\alpha_2}{2}	 & \cos \frac{j\alpha_2}{2}
\end{array} \right),
\label{g5}
\end{equation}  
\begin{equation}
u_3(\alpha_3;j)=e^{\alpha_3T_3(j)}=\left(\begin{array}{cc}
	e^{i\frac{\alpha_3}{2}} & 0 \\
0	 & e^{-i\frac{\alpha_3}{2}}
\end{array} \right).
\label{g6}
\end{equation}  
The corresponding generators 
$$    
  T_1(j)= j\frac{i}{2}\left(\begin{array}{cc}
	0 & 1 \\
	1 & 0
\end{array} \right)=j\frac{i}{2}\tau_1, \quad 
T_2(j)= j\frac{i}{2}\left(\begin{array}{cc}
	0 & -i \\
	i & 0
\end{array} \right)=j\frac{i}{2}\tau_2, %\quad 
$$
\begin{equation} 
T_3(j)= \frac{i}{2}\left(\begin{array}{cc}
	1 & 0 \\
	0 & -1
\end{array} \right)=\frac{i}{2}\tau_3, 
\label{g7}
\end{equation} 
with $\tau_k$ being  Pauli matrices, 
are subject of commutation relations
$$  
[T_1(j),T_2(j)]=-j^2T_3(j), \quad [T_3(j),T_1(j)]=-T_2(j), 
$$
\begin{equation} 
 [T_2(j),T_3(j)]=-T_1(j),
\label{g8}
\end{equation}
 and form the Lie algebra $su(2;j)$ with the general element
\begin{equation} 
 T(j)=\sum_{k=1}^{3}a_kT_k(j)= \frac{i}{2}\left(\begin{array}{cc}
	a_3 & j(a_1-ia_2) \\
	j(a_1+ia_2) & -a_3
\end{array} \right)=-T^{\dagger}(j).
\label{g8-1}
\end{equation}
%Here $\tau_k$ in equations  (\ref{g7}) are Pauli matrices.

There are two more or less equivalent way of group contraction. We can put the contraction parameter equal to the nilpotent unit $j=\iota$  or  tend it to zero $j\rightarrow 0$.  Sometimes it is  convenient to use the first (mathematical) approach, sometimes the second (physical) one. For example, the matrix $u(j)$  (\ref{g3}) has non-zero nilpotent non-diagonal elements for $j=\iota$,
whereas for $j\rightarrow 0$ they are formally equal to zero. Nevertheless both approaches lead to the same final results.
 
Let us describe the contracted group $SU(2;\iota)$ in detail. For $j=\iota$ it follows from (\ref{g3}) that 
$\det u(\iota)=|\alpha|^2=1,$ i.e. $\alpha=e^{i\varphi},$ therefore
\begin{equation} 
u(\iota)= 
\left(\begin{array}{cc}
e^{i\varphi}	 & \iota\beta   \\
-\iota\bar{\beta}	 & e^{-i\varphi}
\end{array} \right), \quad
\beta=\beta_1+i\beta_2 \in {\bf C}. 
\label{g9}
\end{equation}
Functions of nilpotent arguments are defined by their Taylor expansion, in particular, 
$\cos\iota x=1,\;\sin\iota x=\iota x.$ Then one-parameter subgroups of  $SU(2;\iota)$ take the form 
\begin{equation} %$$  
u_1(\alpha_1;\iota)   %=e^{\alpha_1T_1(\iota)}
=\left(\begin{array}{cc}
	1 & \iota i\frac{\alpha_1}{2} \\
	\iota i\frac{\alpha_1}{2} & 1
\end{array} \right), \quad
%$$
%\begin{equation}
u_2(\alpha_2;\iota)   %=e^{\alpha_2T_2(\iota)}
=\left(\begin{array}{cc}
1 & \iota \frac{\alpha_2}{2} \\
-\iota \frac{\alpha_2}{2}	 & 1
\end{array} \right).
\label{g11}
\end{equation}  
The third subgroup does not changed and is given by (\ref{g6}).
The simple group $SU(2)$ is contracted to the non-semisimple  group $SU(2;\iota)$, which is isomorphic to the real  Euclid group $E(2).$
First two generators of the Lie algebra $su(2;\iota)$ are commute $[T_1(\iota),T_2(\iota)]=0$ and the rest commutators are given by (\ref{g8}). For the general element (\ref{g8-1})           %$T(\iota)=a_1T_1(\iota)+a_2T_2(\iota)+a_3T_3(\iota)$ 
of $su(2;\iota)$ the corresponding group element of $SU(2;\iota)$ is as follows 
\begin{equation} 
u(\iota)=e^{T(\iota)}=
\left(\begin{array}{cc}
e^{i\frac{a_3}{2}}	 & \iota i\frac{\bar{a}}{a_3}\sin \frac{a_3}{2}   \\
\iota i\frac{a}{a_3}\sin \frac{a_3}{2} & e^{-i\frac{a_3}{2}}
\end{array} \right), \quad a=a_1+ia_2\in C.
\label{g12}
\end{equation}

The actions of the unitary group $U(1)$ and the electromagnetic subgroup $U(1)_{em}$ 
in the   fibered  space $\Phi_2(\iota)$ are given by the same matrices as on the space $\Phi_2$, namely
\begin{equation}
u(\beta)=e^{\beta Y}=\left(\begin{array}{cc}
	e^{i\frac{\beta}{2}} & 0 \\
0	 & e^{i\frac{\beta}{2}}
\end{array} \right), \quad
u_{em}(\gamma)=e^{\gamma Q}=\left(\begin{array}{cc}
	e^{i\gamma} & 0 \\
0	 & 1
\end{array} \right),
\label{g13}
\end{equation}  
where $Y=\frac{i}{2}{\bf 1}, \; Q=Y+T_3.$
%=\left(\begin{array}{cc}
%	i & 0 \\
%0	 & 0
%\end{array} \right).$

Representations of groups $SU(2;\iota), U(1), U(1)_{em}$   are linear ones,  that is they are realised  by linear operators in  the   fibered  space $\Phi_2(\iota)$.

\section{ Electroweak Model for $SU(2;j)\times U(1)$ gauge group}

The fibered space $\Phi_2(j)$  can be obtained from $\Phi_2$ by 
%substituting $j\phi_2$ instead of $\phi_2.$
substitution $\phi_2 \rightarrow j\phi_2$ in (\ref{g1}),   % or $\chi_2 \rightarrow j\chi_2$ in (\ref{eq11}),
which induces another ones for Lie algebra $su(2)$ generators
$T_1 \rightarrow jT_1,\; T_2 \rightarrow jT_2,\;T_3 \rightarrow T_3. $
As far as the gauge fields take their values in Lie algebra, we can substitute gauge fields instead of transformation of generators, namely
\begin{equation}
A_{\mu}^1 \rightarrow jA_{\mu}^1, \;\; A_{\mu}^2 \rightarrow jA_{\mu}^2,\; \;A_{\mu}^3 \rightarrow A_{\mu}^3, \;\;
B_{\mu} \rightarrow B_{\mu}.
\label{eq28}
\end{equation}  
          % or for the  gauge fields (\ref{eq18})   
          % \begin{equation} %$$
          % W_{\mu}^{\pm} \rightarrow jW_{\mu}^{\pm}, \;\; Z_{\mu} \rightarrow Z_{\mu},\; \;A_{\mu} \rightarrow A_{\mu},
%$$
%\begin{equation}
%{\cal W}^{\pm}_{\mu\nu} \rightarrow j{\cal W}^{\pm}_{\mu\nu},\;\;{\cal Z}_{\mu\nu} \rightarrow {\cal Z}_{\mu\nu},\;\;
%{\cal A}_{\mu\nu} \rightarrow {\cal A}_{\mu\nu}.
          %\label{eq29}
         %\end{equation}
%and the same for  the corresponding abelian stress tensors.  
%The matter field $\phi_1$ does not transformed as well as its small part $\chi$. 
These substitutions in the  Lagrangian $L$ of the %standard 
Higgsless Electroweak Model \cite{Gr-07} %without potential term 
give rise to the  Lagrangian $L(j)$ %(\ref{eq27}).
 %We shall follow the book \cite{R-99} in description of standard Electroweak Model.  
%of the bosonic sector 
of the contracted  model 
with $U(2;j)=SU(2;j)\times U(1)$ gauge group %theory in the space $\Phi_2(j) $  of fundamental representation of $SU(2;j).$ The  bosonic Lagrangian is given by the sum
\begin{equation}
L(j)=L_A(j) + L_{\phi}(j),
\label{eq1}
\end{equation}
where
$$  
 L_A(j)=\frac{1}{2g^2}\mbox{tr}(F_{\mu\nu}(j))^2 + \frac{1}{2g'^2}\mbox{tr}(\hat{B}_{\mu\nu})^2= 
 $$
 \begin{equation}
=  -\frac{1}{4}[j^2(F_{\mu\nu}^1)^2+j^2(F_{\mu\nu}^2)^2+(F_{\mu\nu}^3)^2]-\frac{1}{4}(B_{\mu\nu})^2
\label{eq2}
\end{equation}
is the gauge fields Lagrangian %for $SU(2;j)\times U(1)$ group 
and
\begin{equation}   
  L_{\phi}(j)= \frac{1}{2}(D_\mu \phi(j))^{\dagger}D_\mu \phi(j)            %-V(\phi(j))
\label{eq3}
\end{equation}  
is the {\it free} (without any potential term) matter field Lagrangian (summation on the repeating Greek indexes is always understood).  
%$ \phi(j)= \left(
%\begin{array}{c}
%	j\phi_1 \\
%	\phi_2
%\end{array} \right) \in \Phi_2(j),\;$
Here $D_{\mu}$ are the covariant derivatives
 \begin{equation}
D_\mu\phi(j)=\partial_\mu\phi(j) + g\left(\sum_{k=1}^{3}T_k(j)A^k_\mu \right)\phi(j) + g'YB_\mu\phi(j),
\label{eq4}
\end{equation} 
where $T_k(j)$ are given by (\ref{g7})
 and 
$Y=\frac{i}{2}{\bf 1}$ is generator of $U(1).$ 
Their actions on components of $\phi(j)$ are given by
$$
D_\mu \phi_1=\partial_\mu \phi_1 + \frac{i}{2}(gA_\mu^3+g'B_\mu)\phi_1 + j^2\frac{ig}{2}(A_\mu^1-iA_\mu^2)\phi_2,
$$
\begin{equation}
D_\mu \phi_2=\partial_\mu \phi_2 - \frac{i}{2}(gA_\mu^3-g'B_\mu)\phi_2 + \frac{ig}{2}(A_\mu^1+iA_\mu^2)\phi_1.
\label{eq5}
\end{equation}

The gauge fields 
$$ 
A_\mu (x;j)=g\sum_{k=1}^{3}T_k(j)A^k_\mu (x)=g\frac{i}{2}\left(\begin{array}{cc}
	A^3_\mu  & j(A^1_\mu -iA^2_\mu ) \\
j(A^1_\mu + iA^2_\mu ) & -A^3_\mu 
\end{array} \right),
$$
\begin{equation}
 \hat{B}_\mu (x)=g'YB_\mu (x)=g'\frac{i}{2}\left(\begin{array}{cc}
	B_{\mu} & 0 \\
0 & B_{\mu}
\end{array} \right)
\label{eq6-a}
\end{equation}
 take their values in Lie algebras $su(2;j),$  $u(1)$ respectively,  and the stress tensors are
$$ 
F_{\mu\nu}(x;j)={\cal F}_{\mu\nu}(x;j)+[A_\mu(x;j),A_\nu(x;j)]=
$$
$$
=g\frac{i}{2}\left(\begin{array}{cc}
	F^3_\mu  & j(F^1_\mu -iF^2_\mu ) \\
j(F^1_\mu + iF^2_\mu ) & -F^3_\mu 
\end{array} \right), 
$$
\begin{equation} 
B_{\mu\nu}=\partial_{\mu}B_{\nu}-\partial_{\nu}B_{\mu},         
\label{eq7-a} 
\end{equation}
or in components 
$$
F_{\mu\nu}^1={\cal F}_{\mu\nu}^1  + g(A_\mu^2A_\nu^3-A_\mu^3A_\nu^2), \quad
%$$
%$$
F_{\mu\nu}^2={\cal F}_{\mu\nu}^2 +g(A_\mu^3A_\nu^1-A_\mu^1A_\nu^3),
$$
\begin{equation}
F_{\mu\nu}^3={\cal F}_{\mu\nu}^3 + j^2g(A_\mu^1A_\nu^2-A_\mu^2A_\nu^1),
\label{eq8-a} 
\end{equation}
where ${\cal F}_{\mu\nu}^k=\partial_\mu A_\nu^k-  \partial_\nu A_\mu^k. $

The Lagrangian   $L(j)$ (\ref{eq1}) describe  massless fields. 
In a standard approach to generate   mass terms for the vector bosons the "`sombrero"' potential is added to the matter field Lagrangian 
$  L_{\phi}(j=1)$ (\ref{eq3}) and after that the Higgs mechanism is used.
The different way   %was recently suggested 
\cite{Gr-07}  is based on the fact that the quadratic form $\phi^{\dagger}\phi=\rho^2$
is invariant with respect to gauge transformations. This quadratic form define the 3-dimensional sphere $S_3$ of the radius $\rho>0$ in the target space $\Phi_2$ which is  ${\bf C_2}$ or ${\bf R_4}$ if real components are counted. In  other words  the radial coordinates $R_{+}\times S_3$ are introduced in ${\bf R_4}.$ 
The vector boson masses are  generated by the  transformation of Lagrangian $L(j=1)$ (\ref{eq1}) to the   coordinates on the sphere $S_3$ and are the same as in the standard model. 
Higgs boson field does not appeared if the  sphere radius  does not depend on the space-time  coordinates $\rho=R=const$ \cite{Gr-07}. For $\rho\neq const$ the real positive massless scalar field --- analogy of dilaton or kind of Goldstone mode --- is presented  in the model  \cite{F-08}.

%\section{Higgsless Electroweak Model with 3D spherical matter  space}

The complex  space $\Phi_2(j)$ can be regarded as 4-dimensional real  space ${\bf R}_4(j)$. 
 Let us introduce the real fields    %$r,\; \bar{\psi}(j)=(j\psi_1,j\psi_2,\psi_3)$ by  
\begin{equation}   
\phi_1=r(1+i\psi_3), \quad  \phi_2=r(\psi_2+i\psi_1).   
\label{eq7}
\end{equation} 
The substitution $\phi_2 \rightarrow j\phi_2$  induces the following substitutions 
 \begin{equation}
\psi_1 \rightarrow j\psi_1,\;\psi_2 \rightarrow j\psi_2,\;\psi_3 \rightarrow \psi_3,\; r\rightarrow r
\label{eq7-1}
\end{equation} 
for the real fields.

 For the real fields the form (\ref{g1}) is written as
$r^2(1 + \bar{\psi}^2(j))=R^2, $ where $\bar{ \psi}^2(j)=j^2(\psi_1^2+ \psi_2^2)+\psi_3^2,$ therefore 
\begin{equation}   
r=\frac{R}{\sqrt{1 + \bar{ \psi}^2(j)}}.   
\label{eq9}
\end{equation} 
Hence there are    
three independent real fields $\bar{\psi}(j)=(j\psi_1,j\psi_2,\psi_3).$ These fields  belong to the space $ \Psi_3(j) $
with noneuclidean  geometry which is realized  on the 3-dimensional "`sphere"'  of the radius $R$ in the 4-dimensional  space 
${\bf R}_4(j)$. The fields $\bar{\psi}(j)$ are intrinsic Beltrami coordinates on $ \Psi_3(j)$. The space $ \Psi_3(j=1)\equiv S_3 $
has non degenerate spherical geometry, but  $ \Psi_3(j=\iota) $ is fibered space of constant curvature with 1-dimensional base $\{\psi_3\}$ and 2-dimensional fiber $\{\psi_1,\psi_2\}$ \cite{Gr-09}, so-called semi-spherical space  \cite{P-65}, which can be interpreted as nonrelativistic $(1+2)$ kinematic with curvature or Newton kinematic  \cite{Gr-90}.

%The potential (\ref{eq5}) is the constant  $V(\phi)=\lambda \left(R^2- v^2\right)^2/4  $ 
%on the sphere  and what is more  $V(\phi)=0$ for   $R=v. $ 
The {\it free}  Lagrangian (\ref{eq3}) is transforms to 
     the {\it free} gauge invariant matter field Lagrangian $L_\psi(j) $ on $ \Psi_3(j) $, which is defined with the help of the metric tensor $g_{kl}(j)$ \cite{Gr-07} of  the  space $ \Psi_3(j) $
$$ 
g_{11}=\frac{1+\psi_3^2+j^2\psi_2^2}{(1+\bar{\psi}^2(j))^2}, \quad
g_{22}=\frac{1+\psi_3^2+j^2\psi_1^2}{(1+\bar{\psi}^2(j))^2}, \quad
g_{33}=\frac{1+j^2(\psi_1^2+\psi_2^2)}{(1+\bar{\psi}^2(j))^2}, 
$$
$$
g_{12}=g_{21}=\frac{-j^2\psi_1 \psi_2}{(1+\bar{\psi}^2(j))^2},\;
g_{13}=g_{31}=\frac{-j\psi_1 \psi_3}{(1+\bar{\psi}^2(j))^2},\;
g_{23}=g_{32}=\frac{-j^2\psi_2 \psi_3}{(1+\bar{\psi}^2(j))^2}
$$
 in the form   
$$
L_\psi(j)=\frac{R^2}{2}\sum_{k,l=1}^3g_{kl}(j)D_\mu\psi_k(j)D_\mu\psi_l(j)=
$$
\begin{equation}
=\frac{R^2\left[(1+\bar{\psi}^2(j))(D_{\mu}\bar{\psi}(j))^2-(\bar{\psi}(j),D_{\mu}\bar{\psi}(j))^2\right] }{2(1+\bar{\psi}^2(j))^2}.
\label{eq10}
\end{equation}

The covariant derivatives  (\ref{eq4}) are obtained from the  the representations of generators for the algebras $su(2),$ $u(1)$ in the space $\Psi_3$ \cite{Gr-07} with the help of the substitutions (\ref{eq7-1}) 
$$
T_1\bar{\psi}(j)=\frac{i}{2}\left(\begin{array}{c}
-j(1+j^2\psi_1^2)	 \\
 j(\psi_3-j^2\psi_1\psi_2)  \\
 -j^2(\psi_2+\psi_1\psi_3)
\end{array} \right), \quad
T_2\bar{\psi}(j)=\frac{i}{2}\left(\begin{array}{c}
-j(\psi_3+j^2\psi_1\psi_2) \\
 -j(1+j^2\psi_2^2)\\
 	j^2(\psi_1-\psi_2\psi_3)
\end{array} \right),
$$
$$ 
T_3\bar{\psi}(j) =\frac{i}{2}\left(\begin{array}{c}
j(-\psi_2+\psi_1\psi_3)	\\
j(\psi_1+\psi_2\psi_3) \\
 1+\psi_3^2
\end{array} \right), \quad
Y\bar{\psi}(j) =\frac{i}{2}\left(\begin{array}{c}
-j(\psi_2+\psi_1\psi_3)	\\
 j(\psi_1-\psi_2\psi_3) \\
 -(1+\psi_3^2)
\end{array} \right)
$$
and are as follows: 
$$  %\begin{eqnarray}
D_\mu \psi_1=\partial_\mu \psi_1  - \frac{g'}{2}(\psi_2+\psi_1\psi_3)B_\mu +  
$$
$$
+\frac{g}{2}\left[-(1+j^2\psi_1^2)A_\mu^1 -(\psi_3+j^2\psi_1\psi_2)A_\mu^2-(\psi_2-\psi_1\psi_3)A_\mu^3 \right], 
$$  %\nonumber\\
$$  
  D_\mu \psi_2=\partial_\mu \psi_2  + \frac{g'}{2}(\psi_1-\psi_2\psi_3)B_\mu + 
  $$
  $$
  +\frac{g}{2}\left[(\psi_3-j^2\psi_1\psi_2)A_\mu^1 -(1+j^2\psi_2^2)A_\mu^2 + (\psi_1+\psi_2\psi_3)A_\mu^3 \right], 
$$%\nonumber\\
$$ %\begin{equation}
  D_\mu \psi_3=\partial_\mu \psi_3 - \frac{g'}{2}(1+\psi_3^2)B_\mu +
  $$
 \begin{equation} 
  +\frac{g}{2}\left[-j^2(\psi_2+\psi_1\psi_3)A_\mu^1 +j^2(\psi_1-\psi_2\psi_3)A_\mu^2+(1+\psi_3^2)A_\mu^3 \right]. 
\label{eq11}
\end{equation}  
The gauge fields Lagrangian (\ref{eq2}) does not depend on the fields $\phi$ and therefore remains unchanged.
So the full Lagrangian (\ref{eq1}) is given by the sum of (\ref{eq2}) and (\ref{eq10})

For  small fields, the second order part of the   Lagrangian (\ref{eq10}) is written as
\begin{equation}  
L_\psi^{(2)}(j)=\frac{R^2}{2}\left[(D_\mu \bar{\psi}(j))^{(1)}\right]^2 = 
\frac{R^2}{2}\sum_{k=1}^3\left[(D_\mu \psi_k(j))^{(1)}\right]^2 , 
\label{eq12}
\end{equation} 
where linear terms in covariant derivates (\ref{eq11}) have the form %are 
$$
(D_\mu \psi_1)^{(1)}= \partial_\mu\psi_1-\frac{g}{2}A_\mu^1= -\frac{g}{2}\left(A_\mu^1-\frac{2}{g}\partial_\mu\psi_1\right) =-\frac{g}{2}\hat{A}_\mu^1,
$$
$$  
(D_\mu \psi_2)^{(1)}= \partial_\mu\psi_2-\frac{g}{2}A_\mu^2= 
-\frac{g}{2}\left(A_\mu^2-\frac{2}{g}\partial_\mu\psi_2\right)= -\frac{g}{2}\hat{A}_\mu^2, %\quad
$$
$$ 
(D_\mu \psi_3)^{(1)}=\partial_\mu\psi_3+\frac{g}{2}A_\mu^3-\frac{g'}{2}B_\mu=
\frac{1}{2}\sqrt{g^2+g'^2}Z_\mu.
$$
The new fields 
$$ 
W^{\pm}_\mu = \frac{1}{\sqrt{2}}\left(\hat{A}^1_\mu \mp i \hat{A}^2_\mu \right), \quad %(W^{-}_\mu)^*=W^{+}_\mu,
%$$
%$$ 
 Z_\mu =\frac{gA^3_\mu-g'B_\mu + 2\partial_\mu \psi_3}{\sqrt{g^2+g'^2}}, \quad 
 A_\mu =\frac{g'A^3_\mu+gB_\mu}{\sqrt{g^2+g'^2}}
$$
are transformed as 
$$
W^{\pm}_\mu \rightarrow j W^{\pm}_\mu,\; Z_\mu  \rightarrow Z_\mu,\; A_\mu \rightarrow  A_\mu  
$$ 
and Lagrangian (\ref{eq12}) is rewritten as follows
$$ 
   L_{\psi}^{(2)}= %-\frac{1}{4}\sum_{k=1}^3 ({\cal F}^k_{\mu\nu})^2 - \frac{1}{4}(B_{\mu\nu})^2
  j^2 \frac{R^2g^2}{4}W^{+}_\mu W^{-}_\mu +\frac{R^2(g^2+g'^2)}{8}\left(Z_\mu \right)^2.
$$

The quadratic  part of the full Lagrangian 
$$   
L_0(j)= L_A^{(2)}(j) + L_{\psi}^{(2)}(j)= 
$$
$$ 
=  - \frac{1}{4}({\cal F}_{\mu\nu})^2 -\frac{1}{4}({\cal Z}_{\mu\nu})^2 +\frac{m_Z^2}{2}\left(Z_\mu \right)^2  +
j^2\left\{ -\frac{1}{2}{\cal W}^{+}_{\mu\nu}{\cal W}^{-}_{\mu\nu} + m_{W}^2W^{+}_\mu W^{-}_\mu \right\} \equiv
$$ 
\begin{equation} 
\equiv L_b + j^2 L_f,
\label{neq1}
\end{equation} 
where 
\begin{equation} 
m_W=\frac{Rg}{2}, \quad m_Z=\frac{R}{2}\sqrt{g^2+g'^2},
\label{eq13}
\end{equation}
and  
$
{\cal Z}_{\mu\nu}=\partial_\mu Z_\nu-\partial_\nu Z_\mu, \; 
{\cal F}_{\mu\nu}=\partial_\mu A_\nu-\partial_\nu A_\mu, \; 
{\cal W^{\pm}}_{\mu\nu}=\partial_\mu W^{\pm}_\nu-\partial_\nu W^{\pm}_\mu \; 
$  
are abelian stress tensors  
 describes all the experimentally verified parts of the standard Electroweek Model 
 but does not include the scalar Higgs field. 
% For $R=v$ the masses (\ref{eq13})  are identical to those of the standard Electroweak Model.   %(\ref{eq6}). 
  
The   interaction part of the full Lagrangian in the first degree of approximation is given by
$$ 
L_{int}^{(1)}(j)=j^2\left[L_{A}^{(3)} + L_{\psi}^{(3)}\right],
$$
where the third order terms of the gauge field Lagrangian (\ref{eq2}) are
\begin{eqnarray*}
 L_A^{(3)} =-\frac{g}{\sqrt{g^2+g'^2}}\left\{i\left( \mathcal{W}_{\mu\nu}^-W_\mu^+ - \mathcal{W}_{\mu\nu}^+W_\mu^- \right)\left(g'A_\nu+gZ_\nu \right) -  \right. \nonumber\\
-\frac{\sqrt{2}}{g}\left(g'A_\mu+gZ_\mu \right)    
 \left[\mathcal{W}_{\mu\nu}^+(\partial_{\nu}\psi_2-i\partial_{\nu}\psi_1) + \mathcal{W}_{\mu\nu}^- (\partial_{\nu}\psi_2+i\partial_{\nu}\psi_1) \right]-  \nonumber\\
-\frac{2ig}{\sqrt{g^2+g'^2}}\left(\mathcal{W}_{\mu\nu}^-W_\mu^+ - \mathcal{W}_{\mu\nu}^+W_\mu^- \right)\partial_{\nu}\psi_3 - \nonumber\\
-\frac{2\sqrt{2}}{\sqrt{g^2+g'^2}}\left[\mathcal{W}_{\mu\nu}^+(\partial_{\nu}\psi_2-i\partial_{\nu}\psi_1)  +  \mathcal{W}_{\mu\nu}^-(\partial_{\nu}\psi_2+i\partial_{\nu}\psi_1) \right]\partial_{\nu}\psi_3 + \nonumber\\
+\left(g'\mathcal{F}_{\mu\nu}+ g\mathcal{Z}_{\mu\nu}\right)
  \left\{\frac{i}{4}\left[\left(W_{\mu}^+ \right)^2 - \left(W_{\mu}^- \right)^2 \right]  +
\frac{4}{g^2}\partial_{\mu}\psi_1\partial_{\nu}\psi_2 +  \right. \nonumber\\
\left. \left. 
+ \frac{\sqrt{2}}{g}\left[W_\mu^+ (\partial_{\nu}\psi_2-i\partial_{\nu}\psi_1) + W_\mu^- (\partial_{\nu}\psi_2+i\partial_{\nu}\psi_1) \right]  \right\} \right\} \nonumber\\
\end{eqnarray*}
and those of the matter field Lagrangian (\ref{eq10}) are
\begin{eqnarray*}
 L_\psi^{(3)}= \frac{R^2g}{2\sqrt{2}}\left\{
W_{\mu}^+\left[\psi_3(\partial_\mu \psi_2 - i\partial_\mu \psi_1) -
\frac{g^2-g'^2}{g^2+g'^2}(\psi_2 -i\psi_1) \partial_\mu \psi_3 + \right. \right. \nonumber\\
\left.  + \frac{g'\left(gA_{\mu}-g'Z_{\mu} \right)}{\sqrt{g^2+g'^2}}(\psi_2 -i\psi_1) \right] 
+W_{\mu}^-\left[\psi_3(\partial_\mu \psi_2 + i\partial_\mu \psi_1) - \right.\nonumber\\
\left.  - \frac{g^2-g'^2}{g^2+g'^2}(\psi_2 +i\psi_1) \partial_\mu \psi_3 +  
 \frac{g'\left(gA_{\mu}-g'Z_{\mu} \right)}{\sqrt{g^2+g'^2}}(\psi_2 +i\psi_1) \right]+ \nonumber\\
%$$
%$$ 
\left.+\frac{1}{g}\sqrt{g^2+g'^2}Z_{\mu}\left(\psi_1\partial_\mu \psi_2 - \psi_2\partial_\mu \psi_1 \right)
\right\}. 
\end{eqnarray*}

The fermion Lagrangian of the standard Electroweek Model
is taken in the form \cite{R-99}
\begin{equation} 
L_F=L_l^{\dagger}i\tilde{\tau}_{\mu}D_{\mu}L_l + e_r^{\dagger}i\tau_{\mu}D_{\mu}e_r -
h_e[e_r^{\dagger}(\phi^{\dagger}L_l) +(L_l^{\dagger}\phi)e_r],
\label{eq14}
\end{equation}
where
$
L_l= \left(
\begin{array}{c}
	e_l\\
	\nu_{e,l}
\end{array} \right)
$
is the $SU(2)$-doublet,  $e_r $ the $SU(2)$-singlet, $h_e$ is constant and $e_r, e_l, \nu_e $ are two component Lorentzian spinors. 
Here $\tau_{\mu}$ are Pauli matricies, 
$\tau_{0}=\tilde{\tau_0}={\bf 1},$ $\tilde{\tau_k}=-\tau_k. $ 
The covariant derivatives $D_{\mu}L_l $ are given by (\ref{eq4}) with $L_l$ instead of 
$\phi$ and $D_{\mu}e_r=(\partial_\mu + ig'B_\mu)e_r. $
 The convolution on the inner indices of $SU(2)$-doublet is denoted by $(\phi^{\dagger}L_l)$.

The matter field $\phi$  appears in Lagrangian (\ref{eq14}) only in mass terms. 
When the gauge group $SU(2)$ is contracted to $SU(2;j)$ and the matter field is fibered to $\phi(j)$
the same  take place with doublet $L_l$, namely,  the first component $	e_l$ does not changed, but the second component is multiplied by contraction parameter: $\nu_{e,l} \rightarrow j\nu_{e,l}$. 
With the use of  (\ref{eq7}),(\ref{eq9}) and these substitution the mass terms are rewritten in the form
$$
h_e[e_r^{\dagger}(\phi^{\dagger}(j)L_l(j)) +(L_l^{\dagger}(j)\phi(j))e_r]=
\frac{h_eR}{\sqrt{1+\bar{\psi}^2(j)}}\left\{e_r^{\dagger}e_l + e_l^{\dagger}e_r + \right.
$$
\begin{equation} 
\left.  +i\psi_3\left(e_l^{\dagger}e_r - e_r^{\dagger}e_l \right) + 
ij^2 \left[\psi_1\left(\nu_{e,l}^{\dagger}e_r - e_r^{\dagger}\nu_{e,l} \right)+
i\psi_2\left(\nu_{e,l}^{\dagger}e_r + e_r^{\dagger}\nu_{e,l} \right)\right]
\right\},
\label{n1-1}
\end{equation}
where %$L_l^{\dagger}(j)=(	e_l^{-},j\nu_{e,l}) $ and 
the $SU(2)$-singlet $e_r $ does not transforms under contraction.

\section{Limiting case  of  Higgsless Electroweak Model}

As it was mentioned the vector boson masses are automatically (without any Higgs mechanism) generated by the  transformation of the free Lagrangian  of the standard Electroweak Model  to the Lagrangian (\ref{eq1}),(\ref{eq2}),(\ref{eq10}) expressed in some coordinates on the sphere $\Psi_3(j)$. 
And this statement is true for both values of contraction parameter $j=1,\iota$.
When contraction parameter tends to zero $j^2\rightarrow 0$, then the contribution  of W-bosons fields to the quadratic part of the Lagrangian (\ref{neq1}) will be small in comparison  with the contribution  of Z-boson and electromagnetic fields. 
%The  term $L_h$ being fourth order in $j$ can be neglected.
In other words the limit Lagrangian  includes only   Z-boson and electromagnetic fields. Therefore charded W-bosons fields does not effect on these fields.
% (at least in second order approximation). 
The  part $L_f$ form a new Lagrangian for W-bosons fields and their interactions with other fields. 
The appearance of two Lagrangians $L_b$ and $L_f$ for the limit model is in correspondence with two hermitian forms of fibered  space $\Phi_2(\iota)$, which are invariant under the action of  contracted gauge group $SU(2;\iota)$. 
 Electromagnetic  and Z-boson fields can be regarded  as external fields with respect to the W-bosons fields. 

In mathematical language   the  field space $\left\{ A_{\mu}, Z_{\mu}, W_{\mu}^{\pm}\right\}$
is fibered after contraction $j=\iota$ to the base $\left\{ A_{\mu}, Z_{\mu}\right\}$ and the fiber 
$\left\{W_{\mu}^{\pm}\right\}.$ 
(In order to avoid terminological misunderstanding let us stress that we  have in view locally trivial fibering, which
is defined by the projection $pr:\; \left\{ A_{\mu}, Z_{\mu}, W_{\mu}^{\pm}\right\} \rightarrow \left\{ A_{\mu}, Z_{\mu}\right\}$ in the field space. 
This fibering is understood in the context of  semi-Riemannian geometry \cite{Gr-09} and has nothing to do with the principal fiber bundle.)
Then $L_b$ in (\ref{neq1}) presents Lagrangian in the base and $L_f$ is Lagrangian in the fiber.  In general, properties of a fiber are depend on a points of a base and not the contrary.
In this sense fields in the base are external one with respect to fields in the fiber. 

The fermion Lagrangian (\ref{eq14})
for nilpotent value of the contraction parameter $j=\iota $ is also splited on electron part in the base and neutrino part in the fiber. 
This means that in the limit model electron field is external one relative to neutrino field.
The mass terms (\ref{n1-1}) for $j=\iota $ are
\begin{equation}
h_e[e_r^{\dagger}(\phi^{\dagger}(\iota)L_l(\iota)) +(L_l^{\dagger}(\iota)\phi(\iota))e_r] %=
=\frac{h_eR}{\sqrt{1+\psi_3^2}} \left[ e_r^{\dagger}e_l^{-} + e_l^{- \dagger}e_r  
  +i\psi_3\left(e_l^{- \dagger}e_r - e_r^{\dagger}e_l^{-} \right) \right]. 
\label{n1}
\end{equation}
Its  second order terms 
$ h_eR\left(e_r^{\dagger}e_l^{-} + e_l^{- \dagger}e_r \right)$ 
provide  the electron mass $m_e=h_eR $ and neutrino  remain massless.

Let us note that field interactions in contracted model are more simple as compared with the standard Electroweak Model due to nullification of some terms.
%For example, the last term $j^4L_h$ in (\ref{c1}) disappears as having fourth order in $j\rightarrow 0$. 

\section{Conclusions}

The modified formulation of the Electroweak Model with the gauge group $SU(2)\times U(1)$ based on
the 3-dimensional spherical geometry in the target space is suggested. This model describes all experimentally observed fields and does not include the (up to now unobserved) scalar Higgs field. 
The {\it free} Lagrangian in the spherical matter field space is used instead of Lagrangian  with
the   potential  of the special "`sombrero"' form.
The  gauge field Lagrangian is the standard one. 
There is no need in Higgs   
mechanism since the vector field  masses are generated automatically.

We have discussed the limiting case of the modified Higgsless Electroweak Model, which corresponds to the contracted gauge group $SU(2;j)\times U(1)$, where $j=\iota$ or $j \rightarrow 0$.
The masses of the all experimentally verified particles involved in the Electroweak Model remain the same under contraction, but interactions of the fields are changed in two aspects. 
Firstly  all field interactions become more simpler due to nullification of some terms in Lagrangian.   % are simplyfied. 
Secondly  interrelation  of the fields become more complicated. All fields are divided on two classes: fields in the base
(Z-boson,  electromagnetic and electron) and fields in the fiber (W-bosons and neutrino). 
The base  fields  can be interpreted as  external ones with respect to the fiber fields, i.e.  Z-boson,  electromagnetic and electron fields can interact with W-bosons and neutrino fields, but  W-bosons and neutrino fields do not effect on these fields within framework of the limit model.

This work has been supported in part 
by the Russian Foundation for Basic Research, grant 08-01-90010-Bel-a
and the program "`Fundamental problems of nonlinear dynamics"' of Russian Academy of Sciences.

%\section*{References}

\end{document}